\documentclass[a4paper,10pt]{article}
\usepackage{multicol}
\usepackage{amsmath}
\usepackage{amssymb}
\usepackage{graphicx}
\usepackage{float}
\usepackage{amsmath}
\usepackage{bm}
\usepackage{amsfonts}
\usepackage{balance}
\usepackage{titling}
\usepackage{cite}
\usepackage{color}
\usepackage[title]{appendix}
\makeatletter
\newcommand\figcaption{\def\@captype{figure}\caption}
\makeatother
\newcommand{\bee}{\begin{equation}}
\newcommand{\ee}{\end{equation}}
\newcommand{\beea}{\begin{eqnarray}}
\newcommand{\eea}{\end{eqnarray}}

 \newlength{\halfpagewidth}
 \divide\halfpagewidth by 2

\thanksfootextra{}{)}
\thanksheadextra{}{)}
\thanksmarkseries{alph}
\begin{document}
\title{Collective fermion excitation in a warm massless Dirac system}
\author{ {Daqing Liu$^{1}$\thanks{Corresponding author: liudq@cczu.edu.cn}, \, Shuyue Chen$^{1}$, Shengli Zhang$^2$, Ning Ma$^3$} \\
{\small \it $^{1}$ School of Mathematics and Physics, Changzhou University, Changzhou, 213164, China}\\
{ \small \it $^2$ School of Science, Xi'an Jiaotong University, Xi'an, 710049, China }\\
{\small \it $^3$ Department of Physics, Taiyuan University of Technology, Taiyuan, 030024, China}\\
 }
\date{}
\maketitle

 \begin{abstract}
Basing on a self-consistent method, we predict theoretically that there occurs not only a normal (quasi) fermion mode, but also a collective fermion mode, plasmino, in a warm 2D massless Dirac system, especially in a warm intrinsic graphene system. Results of Landau damping show that both fermion and plasmino are well defined modes. We find that there are sharp differences between the discussed system and the QCD/QED system. Firstly, the thermal mass is proportional to $\alpha_g^{3/4}T$ but not $\alpha_g T$. Secondly, at $0<q<q_c$, the fermion channel and plasmino channel are nearly degenerate and furthermore, the energy difference between fermion and plasmino becomes more and more larger with increasing $q$ at the region $q>q_c$. Thirdly, the fermion behaves as a "relativity particles" with none zero mass and the plasmino exhibits an anormal dispersion at moderate momentum.

 \end{abstract}

{\bf keywords:} plasmino, massless Dirac system, Landau damping.

\begin{multicols}{2}


One of the hot topics of LHC and RHIC is the hot QCD or quark-gluon-plasma (QGP)\cite{qgp}. QGP exists at extremely high temperature and/or high density and is difficult to detect, so an important way to study QGP is to study the collective modes of the system, for instance, collective bosonic mode, plasmon, and collective fermionic mode, plasmino.

Scientists have predicted that at high temperature and/or high density, there are two types of fermionic excitations. One is the well known normal (quasi)fermion branch, and the other is the collective excitation branch, plasmino or antiplasmino\cite{pls1,pls2}. One of the remarkable characteristic of plasmino is that its chirality is opposite to the ordinary (quasi) fermion (In this Letter we only focus on energy larger than zero; we do not distinguish between fermion/plasmino and quasifermion/antiplasmino). The excitation has been extensively investigated in many literatures,
for instance, Refs. \cite{pls3,pls4,pls5,13036698}. Meanwhile, there are still some debates on the mode; for instance, Ref.\cite{jhep,contrav} claimed non-existence of a temperature generated
plasmino mode.

Experimenting on plasmino effects in QCD faces many difficulties. Therefore, one alternative way is to study plasmino in other condensed matter systems, for instance, in superconductors \cite{supercond}. In this Letter we report our study on plasmino in a warm 2D massless Dirac system. We show that, although there are some similarities in the discussed system and QCD system, there are also many striking differences between these systems.

The most famous 2D Dirac system is the graphene system \cite{graphene}, the dispersion of which is $\epsilon_p=\pm v_F p$, where $\epsilon_p$ is the fermion energy with momentum $p$. In this paper we use notations $\hbar=v_F=k_B=1$. In an intrinsic warm massless Dirac system (that is, having no net charge), the fermion propagator with momentum $(p^0,p)$ reads as
\bee
iS_F(p^0,p)=iS_F^0(p^0,p)-2\pi (p^0+\mathbf{\alpha\cdot p}) f_+(p)\delta(p^{02}-p^2),
\ee
where $iS_F^{0} (p)=\frac{i}{2}
[
\frac{1+\alpha\cdot\mathbf{p}/\epsilon_p}{p^0-\epsilon_p+i\eta}+
\frac{1-\alpha\cdot\mathbf{p}/\epsilon_p}{p^0+\epsilon_p-i\eta}
]
=\frac{i(p^0+\alpha\cdot\mathbf{p})}{p^{02}-\epsilon_p^2+i\eta} $ is the fermion propagator without temperature correction and $f_+(p)=\frac{1}{1+e^{p/T}} $  is the so-called  Fermi-Dirac distribution function at temperature $T$.

To obtain the expression of the potential between fermions with momentum $(p^0,p)$, we first notice that in random phase approximation (RPA), the potential takes the form $V(p^0,p)=\frac{V_0(p)}{1+V_0(p)N_c \Pi(p^0,p)}$ where $V_0(p)=2\pi\alpha_g/p$ is the bare Coulomb potential, $\Pi(p^0,p)$ is the polarization and $N_c$ is fermion degenerate ($N_c=4$ in single-layer graphene and $N_c=8$ in double-layer graphene). As pointed by many issues, for instance Eq. (16) in Ref. \cite{coulomb1} or Eq. (38) in Ref. \cite{coulomb2}), in the RPA, we have
\beea
V(p^0,p) &=&\frac{2\pi \alpha_g}{p+\frac{\pi \alpha_g N_c p^2}{8}
(i\frac{\text{sign}p^0\theta(p^{02}-p^2)}{\sqrt{p^{02}-p^2}}+\frac{\theta(p^2-p^{02})}{\sqrt{p^2-p^{02}}})}
\nonumber \\
&\equiv& V^\prime(p_1^0,p_1)/T.
\eea
In the last equation we have used dimensionless quantities $ p_1=p/T$, $p^0_1=p^0/T$ and with the notation $s_d=p^{02}_1-p_1^2$,
\bee
V^\prime(p_1^0,p_1) = \frac{2\pi \alpha_g}{p_1+\frac{\pi \alpha_g N_c p^2_1}{8}
(i\frac{\text{sign}p^0\theta(s_d)}{\sqrt{s_d}}+\frac{\theta(-s_d)}{\sqrt{-s_d}})}.
\ee
If we only consider $p_1^0\gg 1$ and $p\stackrel{<}{\sim}1$, a Taylor expansion according to powers of $p_1/p_1^0$ can be performed on the above equation
\bee
\label{dpot1}
V^\prime(p_1^0,p_1) \simeq \frac{2\pi\alpha_g}{p_1}-\frac{c_2 p_1}{p_1^{02}}-i\frac{c_3}{p_1^0},
\ee
where $c_2=\frac{\pi^3 \alpha_g^3 N_c^2}{32}$ and $c_3=\frac{\pi^2\alpha_g^2N_c}{4}$.


As depicted by Fig. 1, the Dirac fermion self-energy correction is an integral,
\bee
\Sigma(q) = \int \frac{d^3p}{(2\pi)^3} iS_F(p^{0},\mathbf{p})V(p^{\prime 0},\mathbf{p}^\prime) \triangleq \Sigma_r+\Sigma_T
\label{self-energy}
\ee
where $(p^{0\prime},\mathbf{p}^\prime)=(q^0-p^0,\mathbf{q}-\mathbf{p})$ and
\beea
\Sigma_r &=& \int\frac{d^3p}{(2\pi)^3}iS_F^0(p^{0},\mathbf{p})V(p^{\prime 0},\mathbf{p}^\prime), \nonumber \\
\Sigma_T &=& -\int\frac{d^3pf_+(p)}{(2\pi)^2}(p^0+\mathbf{\alpha\cdot p})\delta(p^{02}-p^2) V(p^{\prime 0},\mathbf{p}^\prime). \notag \\
\eea
 As shown in Eq. \ref{self-energy},
the integral can be decomposed into two parts. The first part, $\Sigma_r$, as studied by many literatures (such as Refs. \cite{self1,self2}), corresponds to $v_F$ renormalization at $p^0\simeq v_F p$ and we do not discuss the effect in this paper.

\begin{center}
\begin{minipage}{0.44\textwidth}
\centering
\includegraphics[width=2.6in]{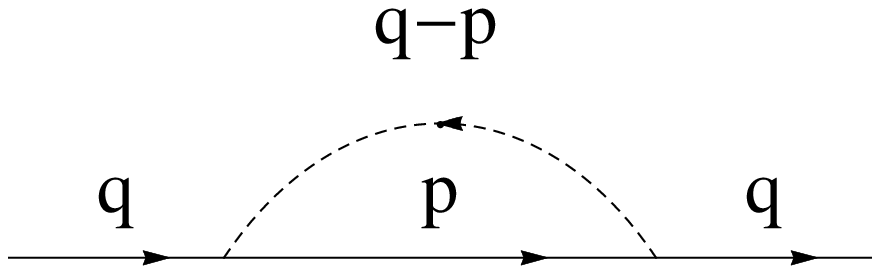}%
\figcaption{The Feynman Diagram of fermion self-energy correction.
}
\end{minipage}
\setlength{\intextsep}{0.in plus 0in minus 0.1in} 
\end{center}

We have
\beea
\Sigma_T &=& \frac{1}{8\pi^2}\int d^2pf_+(p)
[(1-\mathbf{\alpha\cdot \hat{p}})
V(q^0+p,p^\prime)- \nonumber \\ &&
(1+\mathbf{\alpha\cdot \hat{p}})V(q^0-p,p^\prime)],
\eea
where $\hat{\mathbf{p}}=\mathbf{p}/p$.

One can further decompose the above expression into two parts, a scalar part, imaginary part and real one of which are even and odd functions of $q^0$ respectively and can be written as $aq^0$, and a spinor part, which is the even function of $q^0$ and can be written as $-b\alpha\cdot\mathbf{q}$.  With notations of dimensionless quantities $x=q/T$ ($\mathbf{x}=(x,0)$ and $x>0$) $y=q^0/T$, we have,
\beea
a &=& \frac{1}{2\pi^2 y}\int \frac{uv V_-^\prime du dv}{(1+e^u)f_g}, \nonumber \\
b &=& \frac{1}{4\pi^2 x^2}\int \frac{ v(u^2-v^2+x^2)V^\prime_+ du dv}{(1+e^u)f_g},
\nonumber \\ &&
\eea
where $f_g=\sqrt{(u+v)^2-x^2)(x^2-(u-v)^2)}$, $V_\pm^\prime=V^\prime(y+u,v)\pm V^\prime(y-u,v)$ and the domain of  integrations is $u\in(0,\infty),\, v\in(|x-u|,x+u)$.
The factor $1+e^u$ in the denominator of integrands implies that to obtain integrates ($a$ and $b$) we need only focus on $u\ll |y|$ in $V^\prime(y+u,v)$ and $V^\prime(y-u,v)$ (Notice that here we suppose $y\gg 1$). Furthermore, if $x<1$ or $x\sim 1$, the domain of integrations $u\in(0,\infty),\, v\in(|x-u|,x+u)$ implies that it is enough to consider the case of $v\ll |y|-1$ in  $V^\prime(y+u,v)$ and $V^\prime(y-u,v)$. Therefore, to compute $a$ and $b$, taking advantage of the approximate expression of dimensionless potential, Eq. (\ref{dpot1}), is reasonable. Therefore, making Taylor expansion according to powers of $u/y$, one obtains that
\beea
V^\prime_- &\simeq & \frac{4c_2 uv}{y^3}+i\frac{c_3 u}{y^2}
, \nonumber \\
V^\prime_+ &\simeq & \frac{4\pi\alpha_g}{v}-\frac{2c_2 v}{y^2}+f_h(y,u)
,
\eea
where $f_h(y,u)$ is a function independent on $v$.

To obtain approximate expressions of $a$ and $b$, we use identities
\beea
 \int \frac{u^2v dudv}{(1+e^{u})f_g} &=& \frac{3\pi\zeta(3)}{4}, \nonumber \\
 \int \frac{f_h(y,u)v(u^2-v^2+x^2) dudv}{(1+e^{u})f_g} &=&0
\eea
and approximate expressions
\beea
d_1(x) &=& \int \frac{u^2v^2 dudv}{(1+e^{u})f_g}/(\frac{3\pi\zeta(3)}{4}) \nonumber \\
 &\simeq & 3.15+0.13x^2-0.006x^3, \nonumber \\
d_2(x) &=& \int \frac{(u^2-v^2+x^2) dudv}{(1+e^{u})f_g} \nonumber \\
&\simeq & 0.23x^2 \ln(1+\frac{16}{x^3}),\nonumber \\
d_3(x) &=& \int \frac{(u^2-v^2+x^2)v^2 dudv}{\pi (1+e^{u})f_g} \nonumber \\
&\simeq & -0.43x^2 e^{-0.18x}.
\eea
To demonstrate the coincidence of integrals and their approximations, we show them in Fig. 2. It is obvious that in the interesting region all the approximate expression coincide with the corresponding integral very well.

\begin{center}
\begin{minipage}{0.46\textwidth}
\centering
\includegraphics[width=2.6in]{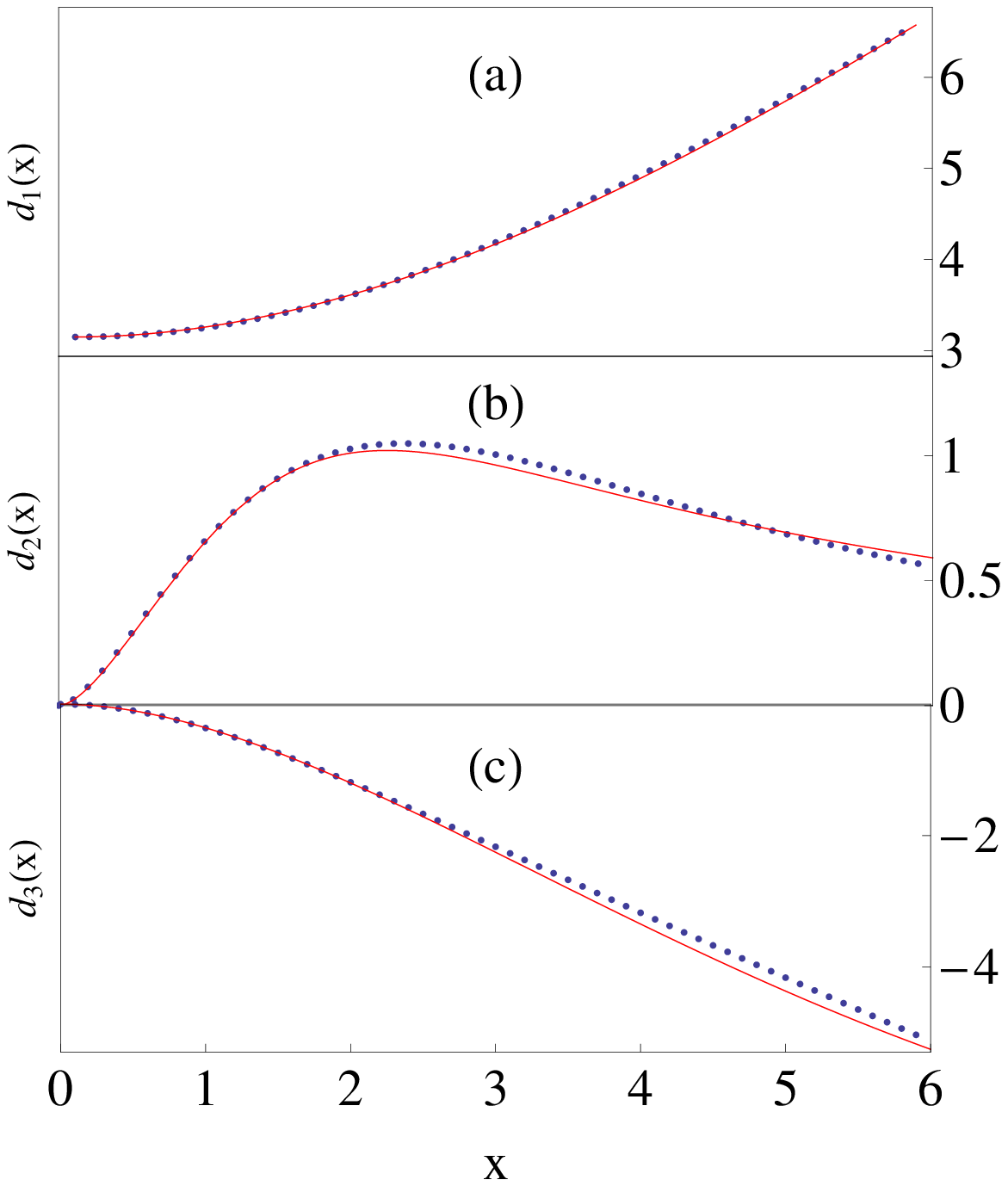}%
\figcaption{Integrals and their corresponding approximate curves.
}
\end{minipage}
\setlength{\intextsep}{0.in plus 0in minus 0.1in} 
\end{center}

Therefore, approximate expressions of $a$ and $b$ are as follows
\beea
a & \simeq & \frac{3\zeta(3)}{8\pi }(\frac{4c_2}{y^4}d_1(x)+i\frac{c_3}{y^3}),
 \notag \\
b &\simeq & \frac{1}{2\pi x^2 }(2\alpha_g d_2(x)-\frac{c_2 d_3(x)}{y^2})
\eea
at $y>0$. Notice that both $\Re a$ and $b$ are even functions of $y$ but $\Im a$, odd function.


In the RPA, the full fermion propagator is,
\bee \label{rpa}
S_{FRPA}^{-1}(q)=S_F^{-1}(q)-\Sigma(q).
\ee
Or, it can be written as
\bee
\tag{\ref{rpa}$'$}
S_{FRPA}^{-1}(q^0,\mathbf{q})=(1-a)q^0-(1-b)\alpha\cdot\mathbf{q}.
\ee
The poles of fermion propagator are twice as many poles as these of the normal zero-temperature fermion propagators, analogous to the results in Refs. \cite{pls1,pls2,pls3}. Noticing that both real parts of $1-a$ and $1-b$ are even functions of $q^0$, one finds that the expression $y=\frac{1-b}{1-a}x$ gives the both the fermion dispersion relation, $\epsilon_p=q^0>0$ and the (anti) plasmino dispersion relation, $\epsilon_h=-q^0>0$.

The fact that $a$ is a complex function ($b$ is a real function) leads to Landau damping both for fermion mode and for plasmino mode. This means that both the  width (or the inverse lifetime) of the fermion mode, $\gamma_p$, and that of the plasmino mode, $\gamma_h$, are nonzero. However, from the expression of $a$, both $\gamma_p/\epsilon_p$ and $\gamma_h/\epsilon_h$, caused by Landau damping, are suppressed by $|T/\epsilon_p|^3$ and $|T/\epsilon_h|^3$. If $\epsilon_p,\epsilon_h>T$ (we shall see in Fig. 4 that this is really the picture), the Landau damping is not important and the excitation modes are well defined. (The energies and widths of the excitation mode can be depicted by the Briet-Wigner approximation). In this case we can at first ignore the imaginary part in $a$ to compute the excitation energies and then replace $y$  by $\epsilon_p$ or $\epsilon_h$ respectively in computing the exciton width in the term $\frac{3\pi \zeta(3) N_c\alpha_g^2}{32y^3}$ (if we set $T=1$).

\begin{center}
\begin{minipage}{0.44\textwidth}
\centering
\includegraphics[width=2.4in]{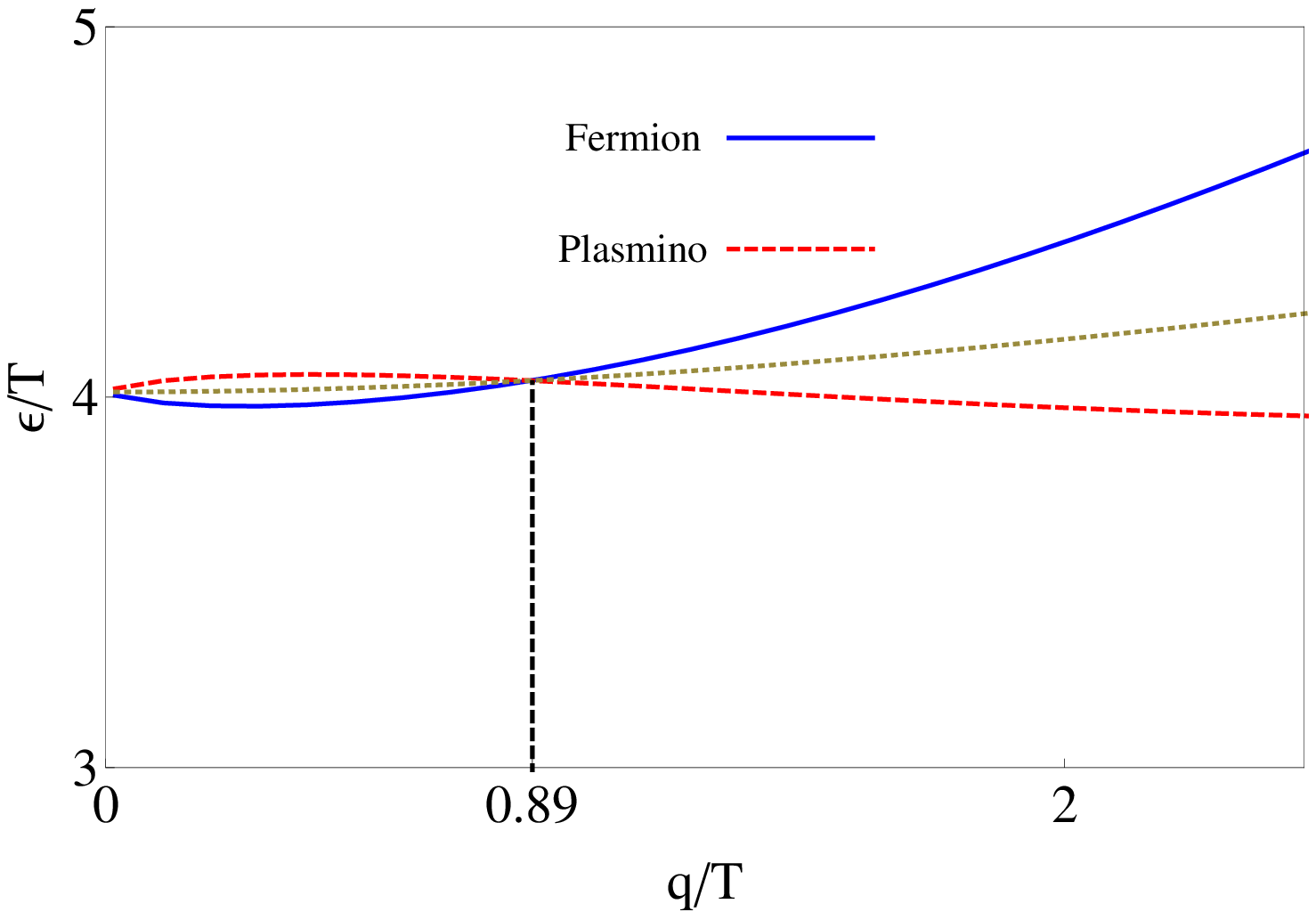}%
\figcaption{Dispersion relations of fermion and plasmino. The solid curve is the dispersion of fermion while the dashed one is the dispersion of antiplasmino. The dotted curve between solid and dashed curves is the expression $0.86N_c^{1/2}\alpha_g^{3/4}d_1^{1/4}(x)$. The vertical line shows the position of $q_c$.
}
\end{minipage}
\setlength{\intextsep}{0.in plus 0in minus 0.1in} 
\end{center}

Choosing a suspended graphene, that is, $\alpha_g=2.1$ and $N_c=4$ , we present in Fig. 3 our results of energy dispersions of fermion excitation and plasmino excitation. (Notice that as pointed out by Ref. \cite{coulomb2}, the predicted semimetal-insulator transition has not yet been observed in experiments in zero magnetic field). We find that there are similarities and several striking differences between the results of our system and the ones of QCD/QED \cite{pls2,pls3,13036698}.

There are some similarities. First, in the larger $x=q/T$ region, the plasmino energy is always below the fermion energy, analogous to QCD and QED \cite{pls1,pls2,13036698}. Second, the plasmino and fermion are the same at $q=0$. In our system the coincident energy at $q=0$ is around $\epsilon_0\simeq 1.15N_c^{1/2}\alpha_g^{3/4}T\simeq 4T$, which is also proportional to temperature $T$.

The most important fact is, however, that our results are significantly different to those of QCD/QED systems. Firstly, for ordinary $N_c=4$ and $\alpha_g\sim 2.1$, {\it i.e.},  $\epsilon_0\propto N_c^{1/2}\alpha_g^{3/4}$ (Note that in a QED/QCD system $\epsilon_0\propto \alpha_g$). Secondly, when $0<q<q_c\simeq 0.89T$, we have an opposite relation between $\epsilon_p$ and $\epsilon_h$, that is, $\epsilon_p<\epsilon_h$. Thirdly, the fermion and antiplasmino have the same energy not only at $q=0$, but also at $q=q_c$, $\epsilon_p(q_c)=\epsilon_h(q_c)\doteq \epsilon_c\simeq 4.04T$. At this point, $1-b(q_c,\epsilon_c)=1-a(q_c,\epsilon_c)= 0$.

In a QCD/QED system, the fermion energy increases monotonically and the collective plasmino mode exhibits a minimum at $q\ne 0$ when momentum $q$ increases. In our system, however, the fermion mode (but not plasmino mode) exhibits as sunken, that is, it has a minimum at $q_1\simeq 0.29T$ and $\epsilon_p(q_1)\equiv m_p\simeq 3.975T$. Furthermore, the behavior of plasmino mode is significantly different from other systems. In the interesting region, it has a maximum at $q_2\simeq 0.42T$ and $\epsilon_h(q_2)\equiv m_h\simeq 4.06T$. At $q>q_2$, $\epsilon_h(q)$ is not a monotonically increasing function but a monotonically decreasing one. This phenomenon may be nominated as plasmino anormal dispersion. To understand the anormal dispersion, we note that, roughly speaking, when the momentum increases, on one hand, the average energy of the trapped particle should generally increase as well, however, on the other hand, the trapping should decrease as the momentum increases. When $q$ is not very large, the trapping decreasing is smaller than the momentum increasing and the plasmino energy is an increasing function. However, and in contrast, when $q>q_2$, the trapping decreasing is larger than the momentum increasing and therefore the plasmino energy is a decreasing function (The point can be seen in Fig. 4).  Since the plasmino is related to the electromagnetic properties of the system, studying the plasmino anormal dispersion and its effect is interesting.

We also list decay widths of fermions and plasminos due to Landau damping in Fig. 4. From the figure one finds that obviously $\gamma_h\ll q\ll \epsilon_h$ at $q<2T$ and $\gamma_p\ll q\ll \epsilon_p$ when $q<3T$. Therefore, both fermion and plasmino are well-defined modes. It is interesting that, approximately, the fermion energies can be approximately depicted by $\sqrt{m_0^2+(q-0.07T)^2}$ at moderate $q$, where $m_0=3.97T$. In other words, the fermion behaves as a "relativity particle" with effective mass $m_0\sim \epsilon_0$. The plasmino energy at moderate $q$ can approximately be depicted by $4T+0.19\sqrt{q T}-0.15 q$.

Fig. 4 shows that $\gamma_p,\gamma_h\sim 0.1T$. Since we have also $m_h-m_p\sim 0.1T$, the tiny rises and falls of the dispersion relation between $q=0$ and $q=q_c$ are at the order of the computed widths of each dispersion. The relative height of energy is not very serious at this region. But, the statement, {\it i.e.}, energies of fermion an plasmino are nearly degenerate in long wave region, is obviously right, which is still sharply different from ordinary QCD/QED system. Since $\epsilon_0\simeq m_p\simeq m_h$, one can nominate $\epsilon_0$ as thermal mass. Notice that we always make Taylor expansion by powers of $q/q_0$ in calculations, our discussions are only valid at $q,T\ll \epsilon_{p(h)}$, which is just the reason that from Fig 4 we restrict our discussions in the region $q<3T$ for fermion channel or in the region $q<2T$ for plasmino channel.

\begin{center}
\begin{minipage}{0.44\textwidth}
\centering
\includegraphics[width=2.7in]{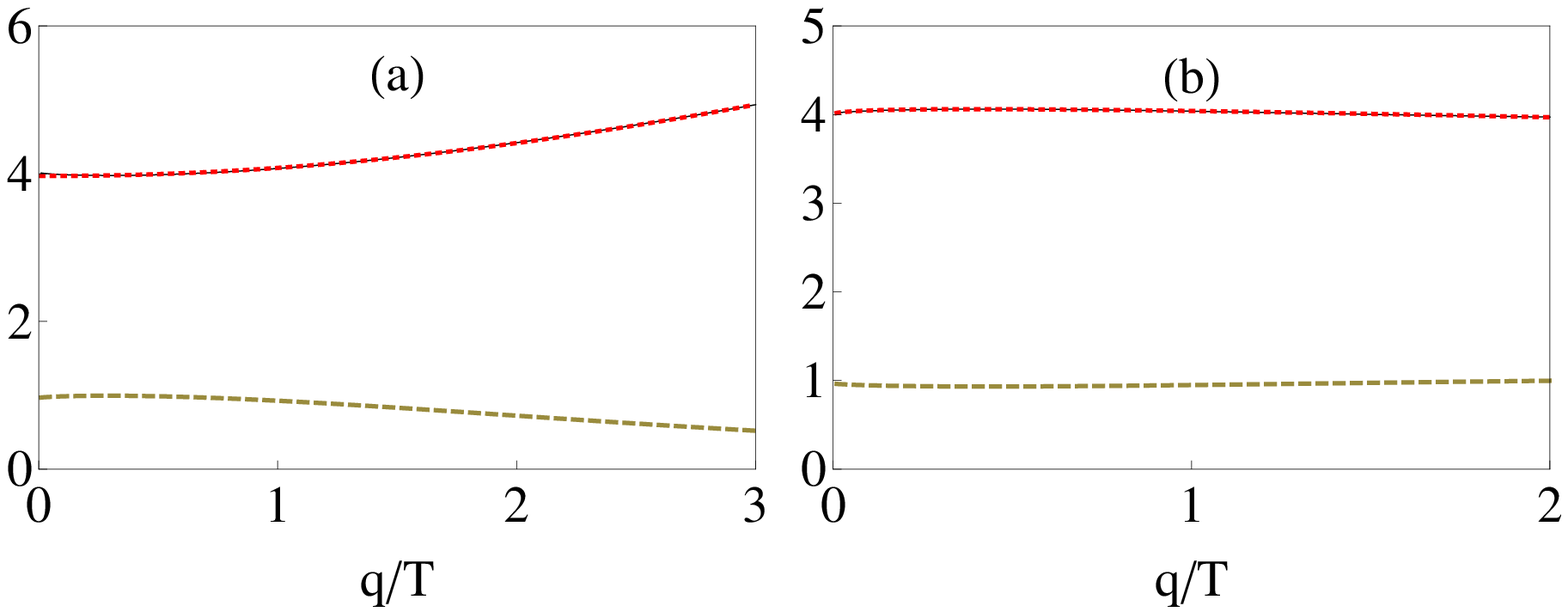}%
\figcaption{Dispersion relations and decay widths of fermion (a) and plasmino (b). Solid curves are energies and dashed ones are decay widths (Here, to clarify, we have multiplied both $\gamma_p$ and $\gamma_h$ by 10). Dotted curves are $\sqrt{m_0^2+(q-0.07T)^2}$ (a) and $4T+0.19\sqrt{q T}-0.15 q$ (b) respectively .
}
\end{minipage}
\setlength{\intextsep}{0.in plus 0in minus 0.1in} 
\end{center}

Some of the interesting properties of the normal and plasmino dispersions in the relativistic case involve the way they approach the light-cone at large momentum, especially in QCD/QED system. It is possibly that we can not discuss these properties in our system, because 1) at large momentum, our discussions are invalid, as pointed out by the above paragraph; 2) the graphene system is only a relativistic-like system but not a strictly relativistic system, for instance, it is hardly to perform a Lorentz transformation in the system. However, if we can make complete calculation, such study is suitable.

In summary, with a self-consistent calculation, at nonzero temperature, we have found that in the long-wavelength region of an intrinsic 2D massless Dirac system there are not only normal collective fermion modes, but also collective plasmino modes, the chiral of which are opposite to the fermion modes. Both energies are on the order of $\epsilon_0\simeq 0.15N_c^{1/2}\alpha_g^{3/4}T\simeq 4T$. Since in the interesting region $\gamma_p\ll \epsilon_p$ and $\gamma_h\ll \epsilon_h$, both fermion and plasmino are well defined modes. However, there are sharp differences between the discussed system and the QCD/QED system. Firstly, $\epsilon_0$ is proportional to $\alpha_g^{3/4}T$ but not the normal one of $\alpha_g T$. Secondly, at $0<q<q_c$, the fermion channel and plasmino channel are nearly degenerate and furthermore, the energy difference between fermion and plasmino becomes more and more large with increasing $q$ at the region $q>q_c$, which significantly deviates from the QCD/QED systems. Thirdly, the mode which has a minimum at $q_1\ne 0$ is not plasmino but fermion; on the contrary, the plasmino has a maximum at $q_2\ne 0$. Although the fermion energy increases monotonically with increasing momentum at $q>q_1$, the plasmino energy decreases monotonically with increasing momentum at $q>q_2$. In this Letter we nominated the interesting phenomenon as anormal dispersion. We believe that our predictions can be tested in a 2D massless Dirac system, specifically, a graphene system, at finite temperature. Note that the material conductivity is related to fermion degree. Our discussions may help to understand the confliction of graphene dc conductivity between experiments and theoretical calculations \cite{con} at $T \to 0$.

The plasmino mode was first predicted in a QCD system. However, the existence of plasmino in QCD is still under debated. We predict that in a 2D massless condensed system, specifically, a graphene system, one can observe the plasmino mode, which is on the order of 0.1eV at room temperature. The prediction can be detected, for instance, by Infrared spectroscopic techniques  on graphene. It is hoped that the results of this study will be helpful in  designing new type of light-emitting devices.

{\bf Acknowledgments}
This work is supported by the National Nature Science
foundation of (Grants No.51176016).


\balance

\end{multicols}

\end{document}